\documentclass[twocolumn,showpacs,amsmath,aps,pra,superscriptaddress]{revtex4}

\usepackage{graphicx}
\usepackage{dcolumn}
\usepackage{bm}

\begin{document}

\title{Quantum diffraction and interference of spatially correlated photon pairs \\ and its Fourier-optical analysis} 

\author{Ryosuke Shimizu}
\affiliation{CREST, Japan Science and Technology Agency, 4-1-8 Honcho, Kawaguchi 332-0012, Japan}

\author{Keiichi Edamatsu}
\affiliation{Research Institute of Electrical Communication, Tohoku University, Sendai 980-8577, Japan}
\affiliation{CREST, Japan Science and Technology Agency, 4-1-8 Honcho, Kawaguchi 332-0012, Japan}

\author{Tadashi Itoh}
\affiliation{Graduate School of Engineering Science, Osaka University, Toyonaka 560-8531, Japan}
\affiliation{CREST, Japan Science and Technology Agency, 4-1-8 Honcho, Kawaguchi 332-0012, Japan}

\date{\hspace*{3cm}}

\begin{abstract}
We present one- and two-photon diffraction and interference experiments involving
parametric down-converted photon pairs. By controlling the divergence of the pump beam in parametric down-conversion, the diffraction-interference pattern produced by an object changes from a quantum (perfectly correlated) case to a classical (uncorrelated) one. The observed diffraction and interference patterns are accurately reproduced by  Fourier-optical analysis taking into account the quantum spatial correlation. We show that the relation between the spatial correlation and the object size plays a crucial role in the formation of both one- and two-photon diffraction-interference patterns.
\end{abstract}


\maketitle

\section{Introduction}
Optical engineering is one of the most precise, sophisticated, and valuable technologies. However, its guiding principle is still based on ``classical wave optics'' \cite{Born}. Recently, the non-classical nature of light has attracted much attention because of its possible application to novel optical technologies, including quantum information processing \cite{Bou00} and quantum imaging \cite{Lugiato02}. Actually, over the past few decades, a considerable number of studies have been conducted on non-classical light, which cannot be explained by classical wave optics. In such studies, photon pairs produced by spontaneous parametric down-conversion have frequently been used because they can be easily generated and can provide a superior environment to realize the concept of the non-classical nature of light. With the development of studies on entangled photons, it has become clear that entanglement, or quantum correlation, between the constituent photons plays a crucial role in the formation of diffraction patterns. Klysyko and his co-workers have carried out pioneering theoretical work in this field \cite{Belinsky94}. Based on this proposal, Strekalov and his co-workers have reported non-classical effects in diffraction phenomena, known as ``ghost-interference'' \cite{Strekalov95}. This work has been developed into one of the fields of quantum imaging. 
Meanwhile, the concept of the ``photonic de Broglie wave'' has been proposed as an intuitive and essential way to understand the interferometric properties of a multiphoton state \cite{Jacobson95}. Within this concept, the photonic de Broglie wavelength of an ensemble photons with wavelength $\lambda$ and number of photons $n$ can be measured to be $\lambda/n$. Recently, a number of experimental demonstrations of this phenomenon have been realized \cite{Fonseca99,Edamatsu01,Shimizu03,Mitchell04,Walther04}. This reduced interferometric property of a multiphoton state has attracted great interest as a light source for novel imaging technology, achieving high-resolution imaging at sub-wavelength resolution. This proposal is known as ``quantum lithography'' \cite{Boto00,Kok01}. D'Angelo and her co-workers have shown, as a proof-of-principle experiment of quantum lithography, that the width of the two-photon diffraction patterns is narrower than that of classical light by a factor of 2 \cite{Angelo01}.
For its simplicity, Young's double slits have been frequently used in these experiments, and the expected counting rate was obtained by calculating a quantum-mechanical fourth-order correlation function. The use of Young's double slit constitutes a simple and thoughtful approach, but the method of directly calculating the fourth-order correlation function is relatively complicated for analyzing arbitrary patterns of two-photon diffraction. 
In this paper, we present a simple method using Fourier-optical analysis for a two-photon state, considering the quantum correlation between the constituent photons. This approach corresponds to the Fraunhofer diffraction of the classical optics, and is applicable to any arbitrary object. To demonstrate the essence of this concept, we measure the diffraction-interference patterns of parametric down-converted photons through a transmission grating, as an example of arbitrary objects, by means of both one- and two-photon detection schemes. We show that the observed patterns can be quantitatively explained by Fourier-optical analysis.

\section{Fourier-optical analysis of a two-photon state}
In the process of parametric down-conversion, the properties (energy, momentum, and their uncertainty) of signal and idler photons are determined by the phase matching condition in a nonlinear crystal, concerning the length and dispersion of the crystal \cite{Burlakov97} and the divergence of pump beam \cite{Grayson94}.
In the following, we assume monochromatic, paraxial, and thin crystal approximation for the sake of simplicity. Under these conditions, we need to consider only the transverse components of the wave vector \cite{Monken98}.
Thus, hereafter, we discuss diffraction-interference in a one-dimensional system.

Although the actual form of a two-photon state can be derived from phase matching condition of parametric down-conversion \cite{Burlakov97, Monken98}, providing an explanation for two-photon diffraction phenomena by an arbitrary two-photon source, we begin to consider the following two-photon state in relation to an object:
\begin{equation}
\left| \psi \right\rangle = \int dx \int dx' F(x,x') \hat{a}^{\dagger}(x) \hat{a}^{\dagger}(x') \left| 0 \right\rangle, \label{b1}
\end{equation}
where $F(x,x')$ represents the two-photon probability amplitude at the object, and $\hat{a}^{\dagger}(x)$ is the creation operator of the photon at a position $x$.
Since the two photons are in the same mode, the two-photon amplitude $F$ at the object should have a symmetrical form. Here, we assume a spatially correlated form of the two-photon amplitude:
\begin{equation}
F(x,x')=\frac{1}{2}\left\{ A(x)B(x') + A(x')B(x) \right\}G(x-x'), \label{b2}
\end{equation}
where $A(x)$ and $B(x')$ are the probability amplitude of photons at the position $x$ and $x'$, respectively, and $G(x-x')$ represents the symmetrical spatial correlation between them.
For instance, photons from the parametric down-conversion process, which is used in our experiment, naturally form such a state, because a signal and its conjugate idler photons generate in almost the same position with some uncertainty in a nonlinear crystal.
The two-photon counting rate $R^{(2)}$ at the object is given by the second-order correlation function:
\begin{equation}
\begin{split}
R^{(2)}(x,x') &= \left\langle \psi \right| \hat{a}^{\dagger}(x') \hat{a}^{\dagger}(x) \hat{a}(x) \hat{a}(x') \left| \psi \right\rangle \\
& = \left| \left\langle 0 \right| \hat{a}(x) \hat{a}(x') \left| \psi \right\rangle  \right|^2\\
& = \left| F(x,x') \right|^2, \label{b10}
\end{split}
\end{equation}
where $\hat{a}(x)$ is an annihilation operator of the photon at the position $x$.
In the same manner, the two-photon counting rate at the far field can be written as a function of the transverse wave number $q$:
\begin{equation}
\begin{split}
R^{(2)}(q,q') &= \left\langle \psi \right| \hat{a}^{\dagger}(q') \hat{a}^{\dagger}(q) \hat{a}(q) \hat{a}(q') \left| \psi \right\rangle \\
&= \left| \left\langle 0 \right| \hat{a}(q) \hat{a}(q') \left| \psi \right\rangle  \right|^2, \label{b3}
\end{split}
\end{equation}
where $\hat{a}^{\dagger}(q)$ and $\hat{a}(q)$ are the creation and annihilation operator of the photon with the transverse wave number $q$, respectively.
The operator $\hat{a}(q)$ at the far field is given by the Fourier-transformation of the operator $\hat{a}(x)$ at the object:
\begin{equation}
\hat{a}(q)= \frac{1}{\sqrt{2\pi}}\int \hat{a}(x) \exp(iqx) dx.
\end{equation}
Thus the two-photon amplitude at the far field is given by
\begin{equation}
\begin{split}
\tilde{F}(q,q') &\equiv \left\langle 0 \right| \hat{a}(q) \hat{a}(q') \left| \psi \right\rangle \\
&= \frac{1}{2\pi}\int dx \int dx' \left\langle 0 \right| \hat{a}(x) \hat{a}(x') \left| \psi \right\rangle \\
& \qquad \times\exp\left[ i(qx+q'x') \right] \\
&= \frac{1}{2\pi}\int dx \int dx'F(x,x') 
\exp\left[ i(qx+q'x') \right]. \label{b4}
\end{split}
\end{equation}
In our case, $A(x)=B(x)$ because the two photons pass through the same object.
Therefore Eq.~(\ref{b4}) is reduced to
\begin{equation}
\begin{split}
\tilde{F}(q,q') &= \frac{1}{2\pi}\int dx \int dx' A(x)A(x')G(x-x') \\
& \quad \times \exp\left[ i(qx+q'x') \right]. \label{b5}
\end{split}
\end{equation}
This expression is exactly the same as Fraunhofer diffraction for a two-photon wave packet except for the introduction of the function of $G(x-x')$.
Unlike other Fourier-optical approaches to two-photon interference \cite{Abouraddy02,Ataure02}, we treat the spatial correlation of the photon pairs as a phenomenological correlation function.
As will be shown later, the transverse correlation function $G(x-x')$ plays a crucial role in the formation of diffraction-interference patterns.
The counting rates for two-photon ($R^{(2)}$) and one-photon ($R^{(1)}$) detection at the object are given by \cite{Burlakov97}
\begin{gather}
R^{(2)}(q,q')=\left|\tilde{F}(q,q')\right|^2,\label{eq4_2}\\
R^{(1)}(q)=\int dq' R^{(2)}(q,q').\label{eq4_3}
\end{gather}

First, we consider two extreme cases in which we can analytically calculate Eqs.~(\ref{eq4_2}) and (\ref{eq4_3}).
One is the case in which the two photons are emitted without any spatial correlation, corresponding to $G(x-x')=\textrm{const}$., i.e., the classical case. The other is the case in which the two photons have a perfect spatial correlation, corresponding to $G(x-x')=\delta(x-x')$. In the former case, there is no transverse correlation between the two photons, and the Eq.~(\ref{b5}) can be rewritten as follows:
\begin{eqnarray}
\tilde{F}(q,q') &=&\frac{1}{2\pi}\int dx A(x) \exp(iqx) \nonumber \\
& & {} \times\int dx' A(x') \exp(iq'x') \nonumber \\
&=&\mathcal{F}[A](q)\mathcal{F}[A](q'),\label{eq4_8}
\end{eqnarray}
where $\mathcal{F}[A]$ represents the Fourier transform of $A$. In this case, the two-photon amplitude is separable into two independent photon amplitudes. Therefore, the two- and one-photon counting rates are given by
\begin{eqnarray}
R^{(2)}(q,q')&\propto&\left| \mathcal{F}[A](q) \right|^2 \left| \mathcal{F}[A](q') \right|^2 \nonumber \\
&\propto& R_{\textrm{cl}}^{(1)}(q) R_{\textrm{cl}}^{(1)}(q'),\label{eq4_9}\\
R_{\textrm{cl}}^{(1)}(q)&\propto&\left| \mathcal{F}[A](q) \right|^2.\label{eq4_10}
\end{eqnarray}
The one-photon counting rate $R_{\textrm{cl}}^{(1)}$ corresponds to the classical Fraunhofer diffraction pattern. In addition, when $q=q'$, the two-photon counting rate is simply the square of the one-photon counting rate,
\begin{equation}
R^{(2)}(q,q)=\left|R_{\textrm{cl}}^{(1)}(q)\right|^2.\label{eq4_11}
\end{equation}

On the other hand, in the case of $G(x-x')=\delta(x-x')$, a pair of signal and idler photons passes together through the same point \textit{x} at the object, and Eq.~(\ref{b5}) can be reduced as follows:
\begin{eqnarray}
\tilde{F}(q,q')&=&\frac{1}{2\pi}\int dx A(x)^2 \exp [i(q+q')x]\nonumber\label{eq4_4} \\
&=&\frac{1}{\sqrt{2\pi}}\mathcal{F}[A^2](q+q'),\label{eq4_4}
\end{eqnarray}
In this case, the two-photon amplitude is not separable into any product of two independent amplitudes. In other words, it is a spatially \textit{entangled} state.
Substituting Eq.~(\ref{eq4_4}) into Eq.~(\ref{eq4_2}), we get the biphoton counting rate:
\begin{equation}
R^{(2)}(q,q')\propto\left| \mathcal{F}[A^2](q+q') \right|^2.\label{eq4_5}
\end{equation}
Although the previous report on two-photon diffraction \cite{Abouraddy01,Gatti03} discussed the case in which photon pairs are detected at separate points $(q \neq q')$, we focus on the case in which photon pairs are detected at the same point $(q = q')$ in terms of high-resolution imaging beyond the classical diffraction limit. Then, the two-photon counting rate can be rewritten as
\begin{equation}
R^{(2)}(q,q)\propto\left| \mathcal{F}[A^2](2q) \right|^2,\label{eq4_6}
\end{equation}
where $2q$ indicates the effective transverse wave number of the biphoton amplitude.
This result indicates that a photon pair behaves as a single particle with the transverse wave number $2q$.
On the other hand, substituting Eq.~(\ref{eq4_5}) into Eq.~(\ref{eq4_3}), the one-photon counting rate becomes constant:
\begin{equation}
R^{(1)}(q)=\textrm{const}.\label{eq4_7}
\end{equation}
This result means that the one-photon diffraction-interference of the biphoton will exhibit no modulation.

From Eqs.~(\ref{eq4_10}) and (\ref{eq4_11}), we understand quite naturally that these results in the case of $G(x-x')=\textrm{const}.$ are compatible with classical optics. However, comparing Eqs.~(\ref{eq4_6}) with (\ref{eq4_11}) and (\ref{eq4_7}) with (\ref{eq4_10}), we see that the results for the biphoton amplitude are quite different from those of two independent photon amplitudes in both one- and two-photon counting rates. Biphotons will exhibit half the modulation period that they would have in the classical case in the diffraction-interference pattern for two-photon detection. Moreover, in one-photon detection, no intensity modulation will be exhibited.
\begin{figure}
	\includegraphics[width=8cm,clip]{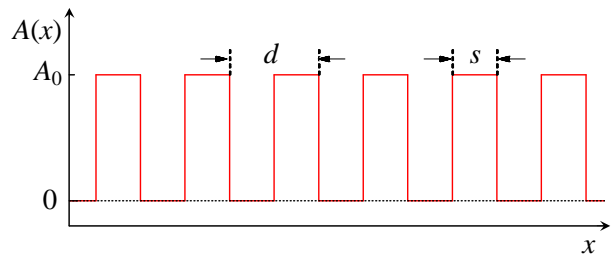}
	\caption{\label{fig:grating}(Color online) Assumed transmission amplitude through the grating, i.e., the grating function, as a function of the transverse position \textit{x}. $A_0$ is the amplitude transmittance through the grating, and \textit{d} and \textit{s} are the slit period and the slit width, respectively. The number of slits was assumed to be \textit{N}.}
\end{figure}

To illustrate the effect of the spatial correlation on the one- and two-photon diffraction-interference, we calculated $R^{(2)}$ and $R^{(1)}$ for some specific examples.
In the following, we assume the rectangular transmission profile of the grating as shown in Figure~\ref{fig:grating}. Note that in this case the Fourier transform of \textit{A} is
\begin{equation}
\mathcal{F}[A](q)=\frac{A_0}{\sqrt{2\pi}}\frac{\sin(Ndq/2)}{\sin(dq/2)}\frac{\sin(sq/2)}{q/2}.\label{eq4_12}
\end{equation}

First, we start with the case of $G(x-x') = \textrm{const}.$, that is, an uncorrelated case.
Substituting Eq.~(\ref{eq4_12}) into Eq.~(\ref{eq4_9}), the two-photon counting rate of two independent photons can be obtained as follows:
\begin{eqnarray}
R^{(2)}(q,q') &=&\left(\frac{A_0}{\sqrt{2\pi}}\right)^4\left| \frac{\sin(Ndq/2)}{\sin(dq/2)}\frac{\sin(sq/2)}{q/2} \right. \nonumber\\
& & {} \times \left. \frac{\sin(Ndq'/2)}{\sin(dq'/2)}\frac{\sin(sq'/2)}{q'/2} \right|^2.\label{eq4_14}
\end{eqnarray}
\begin{figure}
	\includegraphics[width=8.5cm,clip]{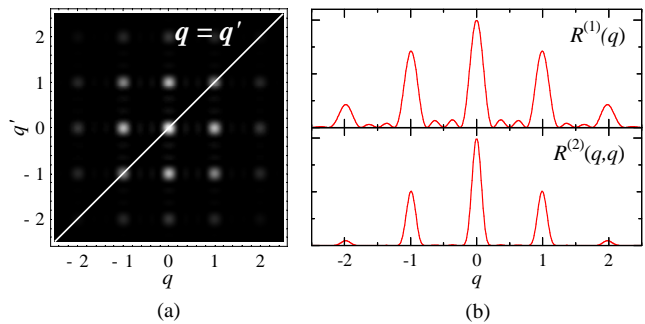}
	\caption{\label{fig:PlotG1}(Color online) (a) Density plot of $R^{(2)}(q,q')$ in the case of $G(x-x')=\textrm{const}$. In this plot, the two-photon counting rate $R^{(2)}(q,q')$ is expressed by a gray scale: the brighter color indicates larger $R^{(2)}$. (b) (Lower graph) Calculated two-photon counting rate $R^{(2)}$ along $q=q'$ . (Upper graph) Corresponding one-photon counting rate $R^{(1)}$.}
\end{figure}
Figure~\ref{fig:PlotG1}(a) shows the density plot of $R^{(2)}(q, q')$ in the case of $G(x-x')=\textrm{const}$. The diffraction transverse wave numbers are normalized by $q = 2\pi/d$, and the ratio $d/s$ is assumed to be 3.2, where $d$ and $s$ are the slit period and the width of the grating, respectively. In this figure, the positions satisfying $q = q'$ are indicated by a solid line. The lower graph in Figure~\ref{fig:PlotG1}(b) denotes the cross-sectional plot of $R^{(2)}$ along $q = q'$. The upper graph in Figure~\ref{fig:PlotG1}(b) represents the one-photon counting rate $R^{(1)}(q)$ obtained from $R^{(2)}(q, q')$ and Eq.~(\ref{eq4_3}). In this case, as expected from Eqs.~(\ref{eq4_8}) - (\ref{eq4_11}), the interference patterns are the same as those obtained by classical optics.

\begin{figure}
    \includegraphics[width=8.5cm,clip]{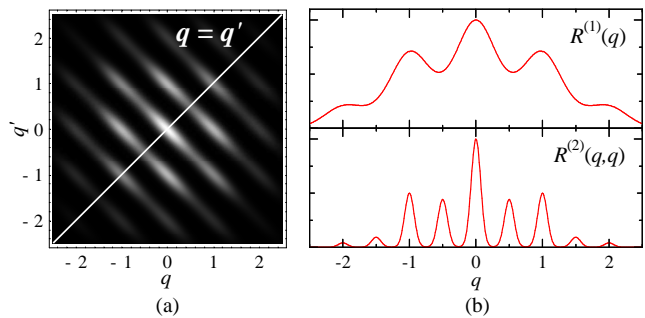}
	\caption{\label{fig:PlotGG}(Color online) (a) Density plot of $R^{(2)}(q,q')$ assuming a Gaussian of $G(x-x')$. (b) (Lower graph) Calculated two-photon counting rate $R^{(2)}$ along $q=q'$ . (Upper graph) Corresponding one-photon counting rate $R^{(1)}$.}
\end{figure}

\begin{figure}
	\includegraphics[width=8.5cm,clip]{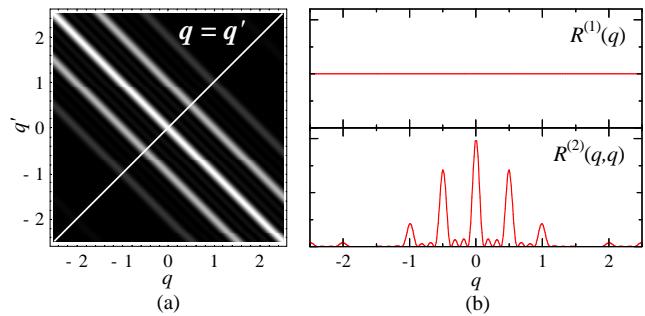}
	\caption{\label{fig:PlotGD}(Color online) (a) Density plot of $R^{(2)}(q,q')$ in the case of $G(x-x')=\delta(x-x')$. (b) (Lower graph) Calculated two-photon counting rate $R^{(2)}$ along $q=q'$ . (Upper graph) Corresponding one-photon counting rate $R^{(1)}$.}
\end{figure}

Next we consider the case of finite spatial correlation. In general, it is difficult to calculate Eq.~(\ref{b5}) analytically. Therefore, we may need to carry out numerical calculations. Here we assume a Gaussian function as the transverse correlation function,
\begin{equation}
G(x-x')=\exp \left[-\left(\frac{x-x'}{(2\sqrt{\ln 2})^{-1} r \, d}\right)^2 \right],
\end{equation}
where $r$ is a dimensionless parameter that expresses the extent of the transverse correlation between the two photons, and corresponds to full width at half maximum (FWHM) of the assumed Gaussian. Figure~\ref{fig:PlotGG}(a) shows the density plot of $R^{(2)}(q, q')$ assuming $r=1$. Comparing Figure~\ref{fig:PlotGG}(a) with Figure~\ref{fig:PlotG1}(a), we understand that the introduction of the transverse spatial correlation $G(x-x')$ results in the convolution of the interference peaks with $g(q-q')$, where $g(q-q')$ corresponds to the Fourier transform of $G(x-x')$. As a result, the peaks of $R^{(2)}$ that are not present in Figure~\ref{fig:PlotG1}(b) appear at $q = q'= \pm0.5, \pm1.5, \ldots,$ as shown in the lower graph of Figure~\ref{fig:PlotGG}(b). In addition, intensity modulations in $R^{(1)}$ disappear as a result. Figure~\ref{fig:PlotGD}(a) presents $R^{(2)}(q,q')$ in the case of $G(x-x') = \delta(x-x')$, which corresponds to perfect spatial correlation. In this case, interference peaks become constant.

\begin{figure}
	\includegraphics[width=8.5cm,clip]{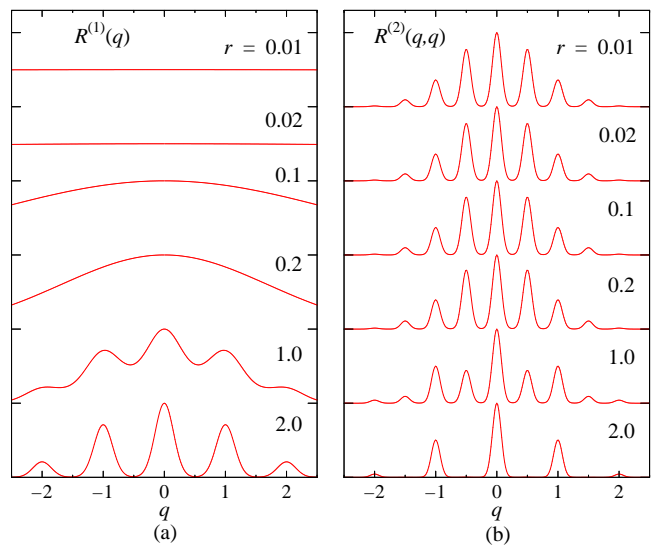}
	\caption{\label{fig:Sim}(Color online) Diffraction-interference patterns with variation of the ratio of distribution of the spatial correlation to the slit period ($r$) in (a) one-photon counting rate $R^{(1)}(q)$ and (b) two-photon counting rate $R^{(2)}(q,q)$.}
\end{figure}

Here we show the numerical calculation results for various values of spatial correlation width $r$. Figure~\ref{fig:Sim} shows the variation of diffraction-interference patterns in both one- and two-photon detection. From these graphs we see that the diffraction-interference patterns in the region in which $r$ is smaller than 0.1 is identical to that in the case of the perfect correlation $\left(G(x-x') = \delta(x-x')\right)$. On the other hand, the diffraction-interference patterns in the region in which $r$ is grater than 2 becomes closer to that in the classical case $(G(x-x') = \textrm{const}.)$. Namely, the relationship between the extent of the transverse correlation and the object size plays a crucial role in forming the diffraction-interference patterns of the biphotons.

\section{Experiment}
\begin{figure}
	\includegraphics[width=8.2cm,clip]{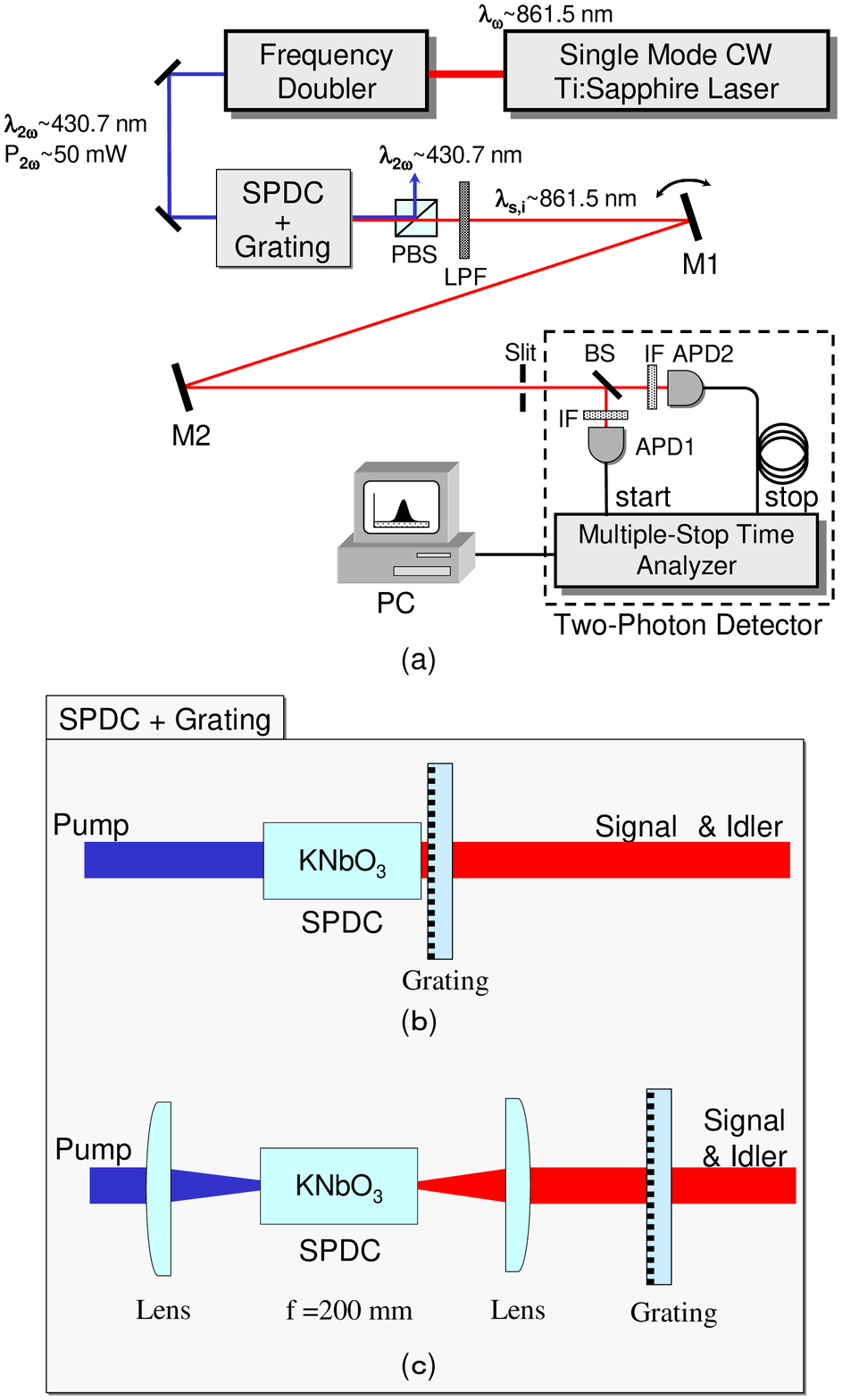}
	\caption{\label{fig:setup}(Color online) (Upper figure) Schematic experimental setup to observe biphoton diffraction and interference. PBS: polarizing beam splitter, LPF: long-pass filter, M1-2: mirrors, BS: non-polarizing beam splitter, IF: interference filters, APD1-2: avalanche photodiodes. (Lower figure) Schematic drawing of the light source used for the biphoton diffraction-interference experiment with (b) perfect and (c) finite spatial correlation. In the case of perfect spatial correlation, we placed the transmission grating just after the KNbO$_3$ crystal. On the other hand, in the case of finite spatial correlation, we put the two lenses (focal length; $\textrm{f}=200$ mm) before and after the KNbO$_3$ crystal in order to increase the spatial correlation width at the grating.}
\end{figure}
Figure~\ref{fig:setup} shows the schematic drawing of the experimental setup. Pairs of frequency-degenerate photons were generated collinearly by spontaneous parametric down-conversion (SPDC) in a 5-mm-long KNbO$_3$ crystal pumped by the second harmonic light (50 mW) of a single longitudinal mode Ti:sapphire laser operating at 861.5 nm. The photon pairs were diffracted by a transmission grating (slit width: 125 $\mu$m, period: 250 $\mu$m). A pair of signal and idler photons has an uncertain emission angle, which originates mainly from the divergence of a pump beam. Therefore, signal and idler photons may pass through the grating slit separately. In order to avoid this case, we placed the transmission grating just after the KNbO$_3$ crystal as shown in Figure~\ref{fig:setup}(b). In this geometry, each photon pair passes together through one of the grating slits \cite{Angelo01}. We separated the photon pairs from the pump beam by using a polarizing beam splitter (PBS) and a long-pass filter (LPF). By rotating a mirror (M1), we recorded spatial diffraction-interference patterns by using a two-photon detector, which consisted of a 50-50 non-polarizing beam splitter (BS) and two avalanche photodiodes (APD; EG\&G SPCM-AQ161) followed by a multiple-stop time analyzer (EG\&G 9308). To measure two-photon coincidence counting rates, we recorded the number of start-stop events within the time window (1 ns). In addition, we simultaneously recorded the number of start pulses as a one-photon counting rate. In front of each APD we placed an interference filter (IF; center wavelength $\lambda _c$ = 860 nm, bandwidth $\Delta\lambda$ = 10 nm). To compare the results with those of classical light, we also observed, using the same apparatus, the diffraction-interference patterns of the Ti:sapphire laser and that of thermal light generated from a tungsten-halogen lamp.

Finally, we observed the diffraction-interference patterns in the case of finite spatial correlation. To achieve this, we observed diffraction-interference by inserting two lenses (focal length; $f = 200$ mm) before and after the KNbO$_3$ crystal and placing the grating at a distance of 40 cm from the crystal as shown in Figure~\ref{fig:setup}(c). The increased divergence of the pump beam enlarges the uncertainty of the transverse wave number of the SPDC photons. This results in the increase of the spatial correlation width between the two photons at the grating \cite{Grayson94}. Therefore, by using the lenses we can control the spatial correlation between the signal and idler photons.

\section{Results and discussion}
\begin{figure}
    \includegraphics[width=7cm,clip]{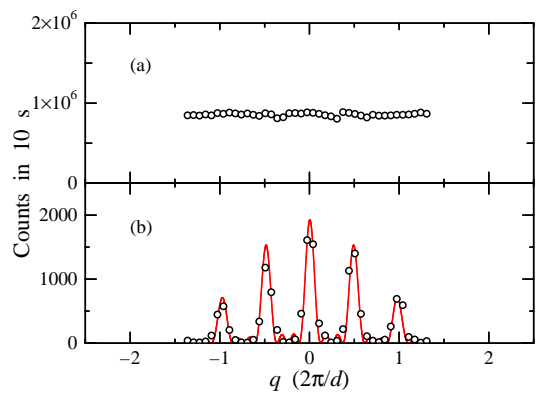}
	\caption{\label{fig:Diff_bd}(Color online) Diffraction-interference patterns of the parametric down-converted photons observed by (a) one-photon detection and (b) two-photon detection.}
\end{figure}
\begin{figure}
	\includegraphics[width=7cm,clip]{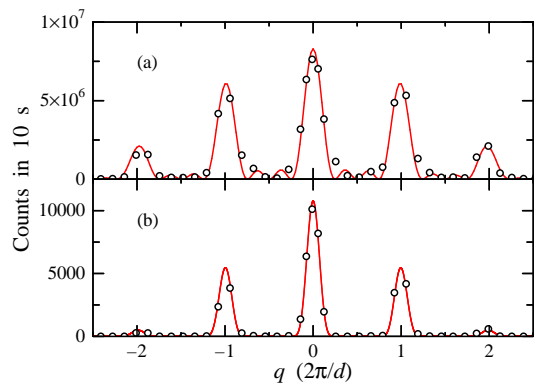}
	\caption{\label{fig:Diff_Ti}(Color online) Diffraction-interference patterns of Ti:sapphire laser observed by (a) one-photon detection and (b) two-photon detection.}
\end{figure}

Figures~\ref{fig:Diff_bd} and \ref{fig:Diff_Ti} show the measured diffraction-interference patterns of parametric down-converted photons and those of the Ti:sapphire laser, respectively. In both figures, the upper graphs represent the interference patterns observed by one-photon detection and the lower graphs are those observed by two-photon detection. The open circles in each figure represent the measured data points. The transverse wave numbers are normalized by $q_0=2\pi/d$, where $d$ is the slit period of the grating.
These experimental data were fitted with the theoretically expected functions (\ref{eq4_10})-(\ref{eq4_11}), (\ref{eq4_6})-(\ref{eq4_7}), as indicated by the solid curves. One can see that the experimental data are in close agreement with the theoretical prediction \cite{Note}. In particular, the two-photon interference of the SPDC seen in Figure~\ref{fig:Diff_bd}(b) exhibits half the modulation period of that of the Ti:sapphire laser in Figure~\ref{fig:Diff_Ti} (see Eq.~(\ref{eq4_6})). This result indicates that the biphoton generated by SPDC behaves as an effective single particle that is associated with half the wavelength of the constituent photons. In other words, the diffraction of a biphoton can be explained by the concept of the photonic de Broglie wave. 

\begin{figure}
	\includegraphics[width=7cm,clip]{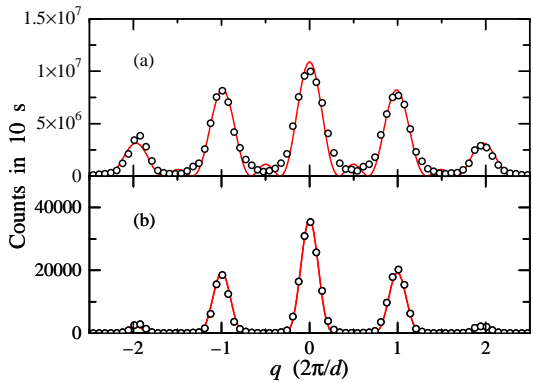}
	\caption{\label{fig:Diff_Ch}(Color online) Diffraction-interference patterns of the halogen lamp observed by (a) one-photon detection and (b) two-photon detection.}
\end{figure}

It is also noteworthy that the one-photon interference of the SPDC exhibits no modulation, whereas that of the Ti:sapphire laser exhibits normal modulation that can be understood by means of classical optics. In classical optics, the disappearance of an interference fringe might be caused by the shortage of coherence length owing to the wide spectral distribution of the parametric emission. To be sure that the observed phenomenon originates from a quantum mechanical effect but not from classical effects caused by the spectral width of the light, we also measured the diffraction-interference pattern of thermal light from a halogen lamp, as shown in Figure~\ref{fig:Diff_Ch}.
Here the thermal light was put through a pinhole with a diameter of 50 $\mu m$ to improve spatial coherence.
We note that the spectrum of the SPDC emission at the detectors is almost the same as that of the halogen lamp, because we detected both emissions through the same interference filters. However, the measured pattern of the halogen lamp is quite different from that of the SPDC, and is quite similar to that of the Ti:sapphire laser. Thus the disappearance of the one-photon diffraction-interference pattern in the SPDC emission is not due to its spectral width.
The disappearance of the one-photon interference pattern is explained by the lack of spatial coherence, which inevitably originates from the spatial correlation of the photons. To discuss the spatial coherence of the constituent photon in a spatially correlated state, we consider the first-order coherence function $g^{(1)}$ defined by
\begin{equation}
g^{(1)}(x,x')=\left\langle \psi \left| \hat{a}^{\dagger}(x) \hat{a}(x') \right| \psi \right\rangle. \label{eq01a}
\end{equation}
We assume that the input state $\left| \psi \right\rangle$ is the two-photon state given in Eq~(\ref{b1}), and the two-photon probability amplitude $F(x, x')$ is given by the following perfectly correlated state:
\begin{equation}
F(x,x')=A(x)A(x')\delta(x-x'). \label{eq01b}
\end{equation}
Substituting Eqs.~(\ref{b1}) and (\ref{eq01b}) into Eq.~(\ref{eq01a}), we get the first-order coherence function:
\begin{equation}
g^{(1)}(x,x')=2\left( \left| A(x) \right|^2 + \left| A(x') \right|^2 \right) A^*(x) A(x') \delta(x-x'). \label{eq01c}
\end{equation}
This result indicates the constituent photons in the spatially correlated state are completely incoherent.
As a result, the one-photon interference pattern is washed-out.
The wide distribution of each transverse wave number $q$ (or $q'$) as shown in Figure~\ref{fig:PlotGD} is another aspect of the spatial incoherence.
Thus the disappearance of the one-photon interference pattern is a natural consequence of the spatial correlation.
\begin{figure}
	\includegraphics[width=7cm,clip]{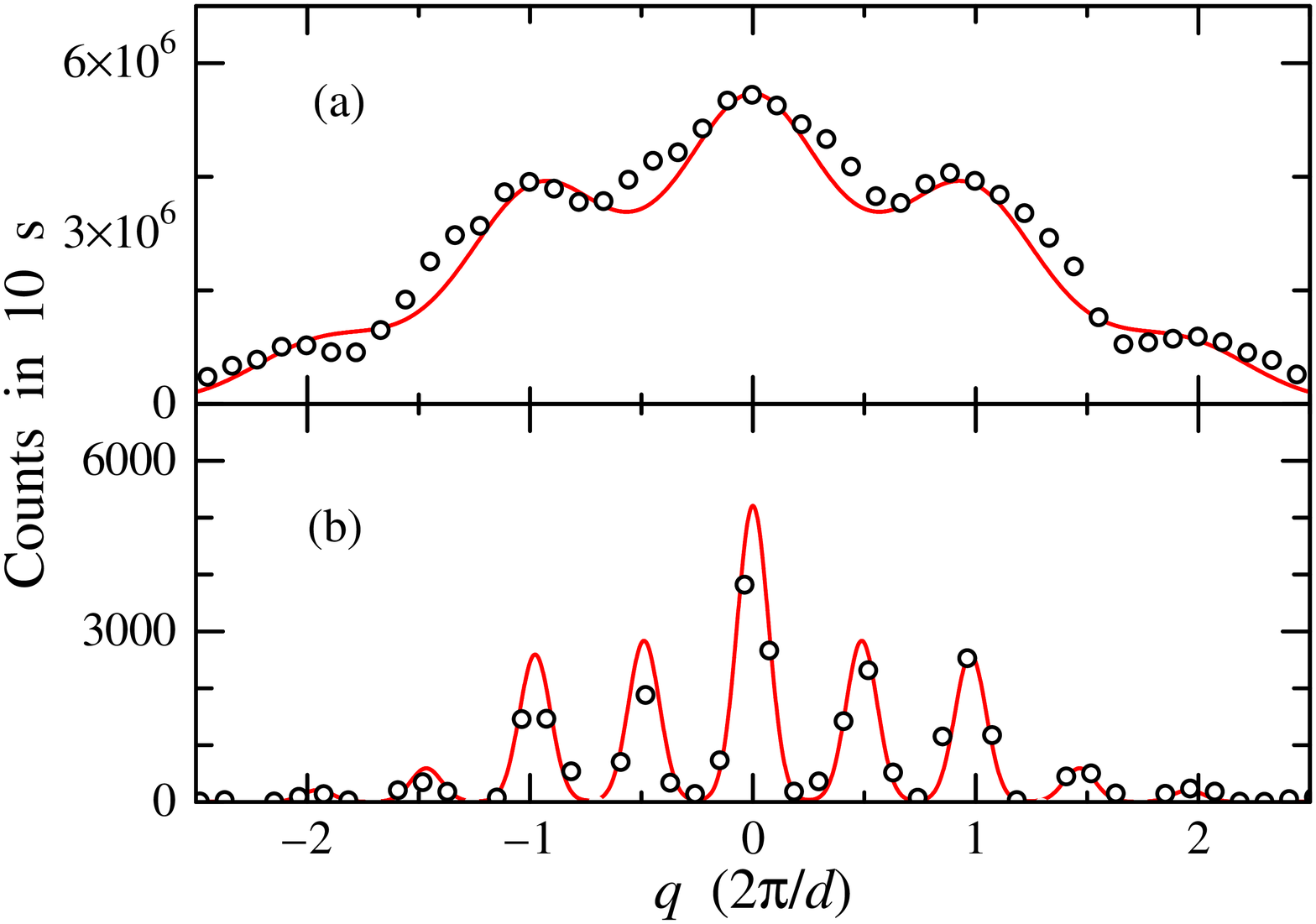}
	\caption{\label{fig:Diff_bm}(Color online) Diffraction-interference patterns of the parametric down-converted photons observed by (a) one-photon detection and (b) two-photon detection when two lenses are inserted before and after the KNbO$_3$ crystal.}
\end{figure}

Next we show the observed diffraction-interference patterns in the case in which the two lenses are placed before and after the KNbO$_3$ crystal, as shown in Figure~\ref{fig:Diff_bm}. Comparing these results with the former ones shown in Figures.~\ref{fig:Diff_bd} and \ref{fig:Diff_Ti}, we see that the diffraction-interference pattern in the two-photon detection method (Fig.~\ref{fig:Diff_bm}(b)) exhibits the intermediate pattern between the two extreme cases. Furthermore, the biphotons begin to show partial modulation in the one-photon detection method (Fig.~\ref{fig:Diff_bm}(a)), again exhibiting the intermediate pattern between Figures~\ref{fig:Diff_bd} and \ref{fig:Diff_Ti}. Considering a finite correlation width, we can also reproduce these patterns from Eqs.~(\ref{b5})-(\ref{eq4_3}). Here we assume the Gaussian correlation function:
\begin{equation}
G(x-x')=\exp \left[-\left(\frac{x-x'}{0.56 d}\right)^2\right],\label{eq16}
\end{equation}
which corresponds to $r=0.78$. 
The calculated patterns are in close agreement with the measured patterns as shown by the solid curves in Figure~\ref{fig:Diff_bm}.
Although some previous experimental reports have focused on spatial correlation in forming the diffraction patterns of a Young's double slit \cite{Riberio94,Abouraddy01A}, we have described the present experiments in terms of high-resolution imaging using spatially correlated photons and also shown the validity of the Fourier-optical analysis taking account of the spatial correlation between constituent photons.
In terms of such high-resolution imaging as quantum lithography, these results indicate that the distribution of a transverse correlation has to be much smaller than an object so as to achieve a spatial resolution high enough to surpass the classical diffraction limit. Otherwise, even entangled photons behave as a classical light source according to the relationship between the extent of the transverse correlation and an object.
However, there is little advantage in using SPDC as an entangled photon source for quantum lithography because the spatial resolution of a pair of parametric down-converted photons is the same as that of the pump beam for SPDC. One of the methods to overcome this problem is to use Hyper Parametric Scattering (HPS) as discussed by Strekalov \cite{Strekalov02}. The spatial resolution of entangled biphotons generated by HPS should be greater than that of the pump beam by a factor of 2, whereas the wavelength of the generated photons is equal to that of the pump beam. By utilizing HPS, in the future it will be possible to obtain a spatial resolution of 100 nm or below by using ultraviolet laser light.
Recently we have successfully generated ultraviolet entangled photons, whose wavelength is approximately 390 nm, via HPS in a semiconductor crystal \cite{Edamatsu04}.
This may open the way to quantum imaging with truly high resolution beyond the classical limit.

\begin{figure}
	\includegraphics[width=8.5cm,clip]{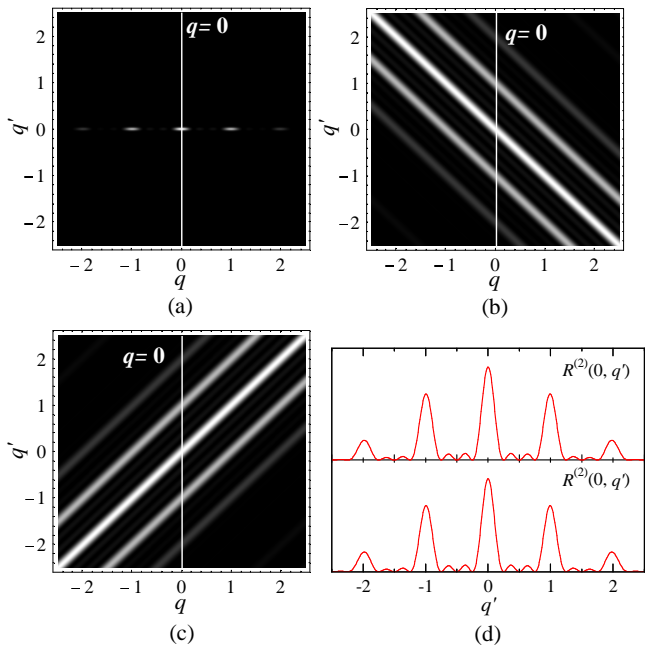}
	\caption{\label{fig:ghost}(Color online) Density plot of $R^{(2)}(q,q')$ in the case of the ghost interference experiment, assuming a grating as the object. (a) Classical case, that is, $G(x-x')=\textrm{const}$. (b) Quantum case, that is, $G(x-x')=\delta(x-x')$. (c) A density plot when the plot (a) oscillates in a direction parallel to the line satisfying $q=q'$. (d) A comparison of the ghost diffraction patterns. The upper graph is the cross-sectional plot along $q=0$ seen in Figure(c) and the lower one is that in Figure(b).}
\end{figure}

Our Fourier-optical analysis is applicable to various types of two-photon diffraction experiment, such as ``ghost interference'' \cite{Strekalov95}. The conditions of the ghost interference experiment differ from those of our experiment in regard to two points.
First, the constituent photons of a photon pair are divided into two arms: the photon amplitude at $q$ (or $q'$) consists of only one photon, whose position is denoted by $x$ ($x'$), scattered by the object $A$ ($B$).
Second, the position of one detector placed after the object is fixed, while the other is scanned.
Under these conditions, the term of $A(x')B(x)$ in Eq.~(\ref{b2}) contributes zero, and thus the two-photon amplitude at the object becomes
\begin{equation}
F(x,x')=\frac{1}{2}A(x)B(x')G(x-x'). \label{c2}
\end{equation}
In addition, the creation and annihilation operators at the object are commutable because the two photons do not exist on the same object.
As a consequence, the two-photon amplitude at the far field is given by
\begin{equation}
\begin{split}
\tilde{F}(q,q') &= \frac{1}{4\pi}\int dx \int dx' A(x)B(x')G(x-x') \\
& \quad \times \exp\left[ i(qx+q'x') \right]. \label{c3}
\end{split}
\end{equation}
Since the object is placed only in one arm, while there is no object in the other arm,
\begin{eqnarray}
\tilde{F}(q,q')&=&\frac{1}{4\pi}\int dx \int dx' A(x) \nonumber \\
& & {} \times G(x-x') \exp [i(qx+q'x')].\label{eq_a00}
\end{eqnarray}
The diffraction pattern is given by $R^{(2)}(0,q')$ in Eq.~(\ref{eq4_2}), which corresponds to the cross-sectional plots along $q=0$.
In the classical case of $G(x-x')=\textrm{const}.$, Eq.~(\ref{eq_a00}) becomes
\begin{eqnarray}
\tilde{F}(q,q')&=&\frac{1}{4\pi}\int dx \int dx' A(x) \exp [i(qx+q'x')] \nonumber \\
&=&\frac{1}{2}\int dx A(x) \exp(iqx) \delta(q').\label{eq_a01}
\end{eqnarray}
Therefore,
\begin{eqnarray}
R^{(2)}(q,q') &\propto& \left| \mathcal{F}[A](q) \delta(q') \right|^2 \nonumber \\
&\propto& \left| R_{\textrm{cl}}^{(1)}(q) \right|^2 \left| \delta(q') \right|^2. \label{eq_q011}
\end{eqnarray}
Note again that $\mathcal{F}[A]$ represents the Fourier transform of $A$.
Thus there is no diffraction pattern as a function of $q'$ (Fig.~\ref{fig:ghost}(a)). On the other hand, in the quantum case of $G(x-x')=\delta(x-x')$,
Eq.~(\ref{eq_a00}) reduces to
\begin{eqnarray}
\tilde{F}(q,q')&=& \frac{1}{4\pi}\int dx A(x) \exp [i(q+q')x] \nonumber \\
&\propto& \mathcal{F}[A](q+q').\label{eq_a02}
\end{eqnarray}
Thus, in accordance with Eqs.~(\ref{eq4_2}), (\ref{eq4_3}) and (\ref{eq4_10}), the two-photon counting rate is given by
\begin{eqnarray}
R^{(2)}(q,q') &\propto& \left| \mathcal{F}[A](q+q') \right|^2 \nonumber \\
&\propto& R_{\textrm{cl}}^{(1)}(q+q').\label{eq_a03}
\end{eqnarray}
This result indicates that the diffraction pattern presented by spatial entangled photon pairs appears as a function of $q'$. As explained in the preceding section, we can also understand this image formation process as the diffraction pattern spread in the direction perpendicular to the line satisfying $q=q'$ as a result of the spatial entanglement (Fig.~\ref{fig:ghost}(b)).

\begin{figure}
	\includegraphics[width=8cm,clip]{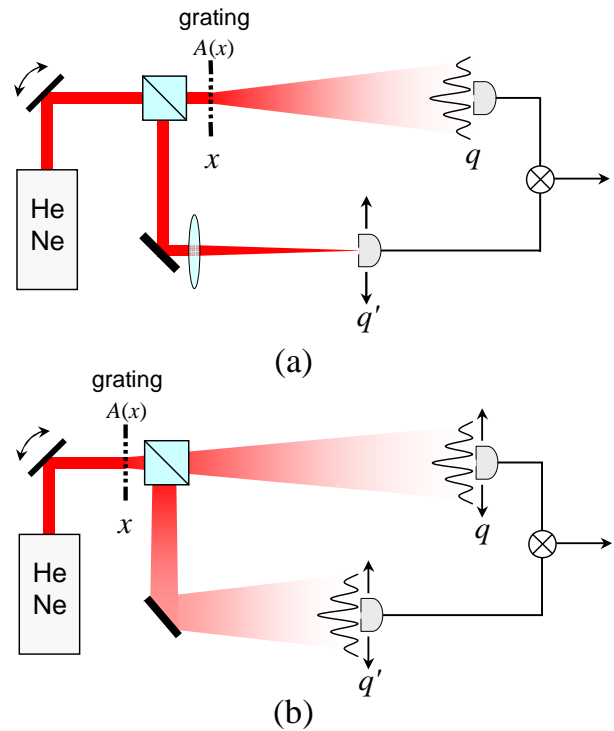}
	\caption{\label{fig:ghostset}(Color online) (a) Schematic view of the ghost interference experiment performed in Ref.~\cite{Bennink04}. Here we put a grating as the object instead of a double slit. (b) An assumed classical version of the experiment described in this paper.}
\end{figure}

\begin{figure}
	\includegraphics[width=8.5cm,clip]{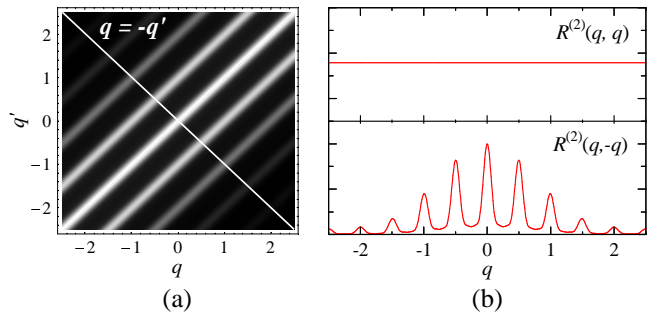}
	\caption{\label{fig:classical}(Color online) (a) A density plot when the $R^{(2)}(q,q')$ in Figure~\ref{fig:PlotG1}(a) oscillates in a direction parallel to the line satisfying $q=q'$. (b) (Upper) A cross-sectional plot along $q=q'$ and (Lower) that along $q=-q'$ in the plot (a).}
\end{figure}

Recently, several groups have reported coincidence imaging experiments using classical light sources, which stimulated discussion regarding the difference between quantum and classical coincidence imaging \cite{Bennink02,Gatti03,Angelo03,Bennink04,Angelo04}. 
Two types of experiment use classical light sources.
One is the experiment using a classically momentum-correlated beam \cite{Bennink02,Bennink04}.
The other is that using thermal light sources \cite{Gatti04,Velencia05,Xiong05,Ferri05}.
Here we discuss only the experiments using a classically momentum-correlated beam in terms of our Fourier-optical analysis because a discussion taking into account a statistical average is required for incoherent illumination, such as in the experiment using thermal light.
The most significant feature of the experiment using momentum-correlated beams is that the diffraction pattern of classically momentum-correlated beams, produced by rotating a mirror before the beam splitter as shown in Figure~\ref{fig:ghostset}(a), appears as a function of $q'$, whereas there is no spatial entanglement on the object, that is, $G(x-x')=\textrm{const}$.
We can understand this image formation process as follows. As shown in Figure~\ref{fig:ghost}(a), the intensity modulation of classical beams appears as a function of $q$ but not $q'$. However, by rotating the mirror, the diffraction pattern shown in Figure~\ref{fig:ghost}(a) oscillates in a direction parallel to the line satisfying $q=q'$ with the scanning frequency of the mirror.
Therefore, taking an average of the product of the two signals over a sufficiently long period compared to the scanning period of the mirror, we can obtain the two-photon counting rate as follows: 
\begin{eqnarray}
R^{(2)}(q,q') &=&\int R^{(1)}_{\textrm{A}}(q-q_{0})R^{(1)}_{\textrm{B}}(q'-q_{0}) dq_{0},\label{eq_a1}
\end{eqnarray}
where $q_{0}$ represents the displacement of the pattern by the rotation of the mirror, and $R^{(1)}_{\textrm{A}}$ and $R^{(1)}_{\textrm{B}}$ are the one-photon counting rate from each object having the form of $A(x)$ and $B(x)$, respectively. Since $R^{(1)}_{\textrm{B}}(q'-q_{0})$ is given by the delta function $\delta(q'-q_{0})$ as a result of $B(x')=\textrm{const}.$ in the ghost diffraction experiment, Eq.~(\ref{eq_a1}) can be reduced as follows:
\begin{eqnarray}
R^{(2)}(q,q') &=&\int R^{(1)}_{\textrm{A}}(q-q_{0})\delta(q'-q_{0}) dq_{0},\nonumber\\
&=& R^{(1)}_{\textrm{A}}(q-q').\label{eq_a2}
\end{eqnarray}
The density plot of Eq.~(\ref{eq_a2}) is shown in Figure~\ref{fig:ghost}(c). This is a symmetric plot of that in Figure~\ref{fig:ghost}(b) with respect to $q=0$. Thus, the cross-sectional plots along $q=0$ seen in Figures~\ref{fig:ghost}(b) and (c) give the same pattern, as shown in Figure~\ref{fig:ghost}(d).
Therefore, to identify the difference between the pattern produced by quantum correlated photons and that by classically correlated beams in the case of ghost interference experiments, more detailed discussion is required as reported in Ref.~\cite{Bennink04,Angelo04}.
On the other hand, the difference clearly appears in the case of the experiment in which two photons pass through the same object, that is, $R^{(1)}_{\textrm{A}}(q)=R^{(1)}_{\textrm{B}}(q)$ in Eq.~(\ref{eq_a1}).
An assumed schematic setup of the classical version of this kind of experiment is shown in Figure~\ref{fig:ghostset}(b), when the form of the transmission amplitude $A(x)$ is the same as that in Figures~\ref{fig:PlotG1}-\ref{fig:PlotGD}.
Figure~\ref{fig:classical}(a) shows the expected density plot, which corresponds to the pattern shown in Figure~\ref{fig:PlotG1}(a), a pattern that oscillates in a direction parallel to the line satisfying $q=q'$. 
In addition, the upper and lower graphs in Figure~\ref{fig:classical}(b) represent their cross-sectional plots along $q=q'$ and $q=-q'$, respectively.
Interference peaks appear with half a modulation period of the normal diffraction-interference pattern shown in Figure~\ref{fig:PlotG1} in the case of $q=q'$, while it shows no modulations in the case of $q=q'$.
To clarify the image formation process, we rewrite Eq.~(\ref{eq_a1}) as follows:
\begin{eqnarray}
R^{(2)}(\Delta)=\int R^{(1)}_{\textrm{A}}(\Delta+Q) R^{(1)}_{\textrm{A}}(Q) dQ,\label{eq_a3}
\end{eqnarray}
where
\begin{eqnarray}
Q=q'-q_{0}, \ \Delta = q-q'.\label{eq_a4}
\end{eqnarray}
If we detected at the same point, i.e., $q=q'$, the two-photon counting rate $R^{(2)}(\Delta)$ becomes constant as a result of $\Delta=0$.
On the other hand, in the case of detection at the symmetrical position $q=-q'$, Eq.~(\ref{eq_a3}) becomes 
\begin{eqnarray}
R^{(2)}(2q)=\int R^{(1)}_{\textrm{A}}(Q+2q) R^{(1)}_{\textrm{A}}(Q) dQ.
\end{eqnarray}
This equation means that the interference of the two photons has half a modulation period of the one-photon counting rate $R^{(1)}_{\textrm{A}}(q)$.
However, the two-photon counting rate $R^{(2)}$ is nothing more than the autocorrelation function of the one-photon counting rate $R^{(1)}_{\textrm{A}}$.
Thus, the diffraction width of this pattern is no narrower than that of the normal patterns as shown in Figure~\ref{fig:PlotG1}(b).
This result is consistent with the consideration that spatial resolution beyond the diffraction limit can be achieved only by using spatially entangled photons but cannot be achieved using a classical light source, as discussed in Ref.~\cite{Angelo03}.

\section{Conclusions}
The experiment described in this paper focuses on the use of entangled photons for novel quantum imaging technologies, which has been attracting much attention in connection with the concept of the photonic de Broglie wave.
We have investigated the role of spatial correlation in diffraction-interference phenomena by measuring the spatial diffraction-interference patterns of spontaneous parametric down-converted biphotons through a transmission grating.
We have shown that the spatially correlated biphoton exhibits half the modulation period of that of classical light. Also, we have confirmed that the one-photon interference of the biphoton exhibits no modulation. Furthermore, by controlling the spatial correlation between the two photons, we have successfully demonstrated that the diffraction-interference pattern changes from the perfectly correlated biphoton case to the classical case. In order to understand these experimental results in a quantitative and straightforward manner, we developed the Fourier-optical analysis of a two-photon state taking into account the spatial correlation between the two photons.

In our experiment, we have investigated the diffraction-interference patterns formed by a one-dimensional regular grating, and demonstrated the validity of Fourier-optical analysis. Nevertheless, we would like to emphasize that this analysis is extendable to arbitrary one- and two-dimensional objects, and can thus be is very useful for future imaging technologies utilizing entangled photons.

\begin{acknowledgments}
We are grateful to H. Ishihara for his valuable discussions.
This work was supported in part by the program ``Strategic Information and Communications R \& D Promotion Scheme'' of the Ministry of Internal Affairs and Communications of Japan.
\end{acknowledgments}


\begin{references}
\bibitem{Note} From the fitting we found that the ratio of the slit period to the slit width (\textit{d/s}) is 3.3$\sim$3.8, whereas it should be 2 for the actual grating we used. This discrepancy is not due to the quantum mechanical effect because the same tendency was observed for both biphotons and classical lights. Although the reason for the discrepancy is not clear at present, it might be due to the longitudinal groove pattern of the grating that was printed on a glass plate.
\end{references}
\end{document}